\documentclass[useAMS,usenatbib]{mn2e}
\usepackage{color}
\usepackage{ulem}
\usepackage{graphicx}

\newcommand{\dn}{$\Delta\nu$}
\newcommand{\ie}{{\it i.e.}}
\newcommand{\eg}{{\it e.g.}}

\title[]{Warping modes in discs around accreting neutron stars}
\author[H. Meheut and  M. Tagger]
{H. Meheut$^{1}$\thanks{E-mail:
hmeheut@apc.univ-paris7.fr} and M. Tagger$^{2, 1}$\\
$^{1}$AstroParticules et Cosmologie (APC), Universit\'e Paris Diderot/CNRS, Paris, France\\
$^{2}$Laboratoire de Physique et Chimie de l'Environnement et de l'Espace (LPC2E), Universit\'e d'Orl\'eans/CNRS, France}
\begin{document}

\date{}


\maketitle

\label{firstpage}

\begin{abstract}
The origin and stability of a thin sheet of plasma in the magnetosphere of an accreting neutron star is investigated. First the radial extension of such a magnetospheric disc is explored. Then a mechanism for magnetospheric accretion is proposed, reconsidering the bending wave explored by \cite{AGA97}, that was found to be stable in ideal MHD. We show that this warping becomes unstable and can reach high amplitudes, in a variant of Pringle's radiation-driven model for the warping of AGN accretion discs (\cite{PRI96}). Finally we discuss how this mechanism might give a clue to explain the observed X-ray kHz QPO of neutron star binaries.
\end{abstract}

\section{Introduction}

Neutron star binaries are at the centre of numerous investigations since they are the laboratory of extreme physics, combining high magnetic fields and strong gravity. These effects dominate the interface between the star magnetosphere and the accretion disc, with which the present work is concerned. Although many works have been dedicated to descriptions of this interface, a fully consistent model is still not available. In the pioneering work of \cite{GL79}, 

a strong anomalous resistivity was assumed in the disc; this allows magnetic field lines anchored in the neutron star and rotating with it to thread the disc over an extended transition region, while the gas remains  in differential rotation. More recent works have rather considered a weaker  resistivity, resulting in a sharp transition between the disc and the magnetosphere. \\
In subsequent work \citet{SPR90} showed that an interchange instability, at the disc/magnetosphere interface, could
allow the gas to penetrate from the keplerian disc into the magnetosphere, and they studied its subsequent fate. In
particular they showed that, depending on the configuration of the magnetosphere, the gas could reach a radius 
where its vertical motion (along the field lines) could in turn become unstable to a warping mode, leading to 
magnetospheric accretion. However \citet{LEP96} later found  that the gas could form a stable magnetospheric disc, 
supported against gravity by the magnetic field in a structure very similar to the `helmets' of stable cold gas 
suspended on magnetic field lines, commonly observed in the solar chromosphere. Furthermore they found that the 
interchange mode was strongly stabilized by line tying, \ie~the fact that, since magnetospheric field lines are 
tied to the surface of the neutron star, they need to be bent somewhere between the surface and the disc to accomodate interchange motion in the disc; the required energy is large and is expected to stabilize the mode in most realistic magnetic configurations.\\
Numerical works \citep{ROM02, ROM03, BZ08} have addressed this question, but they are still dominated by a relatively strong dissipation. Thus, although they can address the extremely complex case where the rotation axis of the disc, the spin axis of the neutron star, and its dipolar magnetic moment are misaligned, and do show the possibility of magnetospheric accretion, this dissipation can be considered as modeling the effect of these instabilities, but they cannot study the instabilities themselves.\\

In these works the emphasis was on possible instabilities of purely MHD origin, for which instability criteria were derived. In this case, when the modes are unstable their frequency is purely imaginary; they have a zero real frequency  ($\rm Real(\nu_{osc})=0$) in the gas frame, and would always be observed Doppler-shifted to rotate at the neutron star rotation frequency $\nu_{*}$.\\ In the present work we reconsider warp oscillations that are stable in the pure MHD context, but become unstable by a different process of radiative origin.\\

We first present in section \ref{magnetospheric} the general setup and geometry
we consider,  and we show how gas can enter the neutron star magnetosphere to
form a disc. In section \ref{warping} we show that the bending mode considered
by \cite{AGA97} is stable in the configuration considered here but that
radiative forcing can make it unstable, in a variation of the mechanism
developed by \citet{PRI96} to explain the warping of the external part of AGN
accretion discs. We also discuss how this instability, since it gives the
gas strong vertical oscillations, might provide the transition from radial
accretion (in the keplerian disc, followed by the interchange mode at its inner
edge) to magnetospheric accretion along magnetic field lines, onto the neutron
star. \\
We note here that our approach is different from that of \cite{FER08} and \cite{OGI01}, who consider warps of the differentially rotating disc (not the magnetospheric disc we consider here), excited by various mechanisms. It is also different from the warps observed in numerical MHD simulations \citep{ROM03} when the stellar magnetic field and the disc plane are misaligned.
\begin{figure}
 \includegraphics[width=0.5\textwidth]{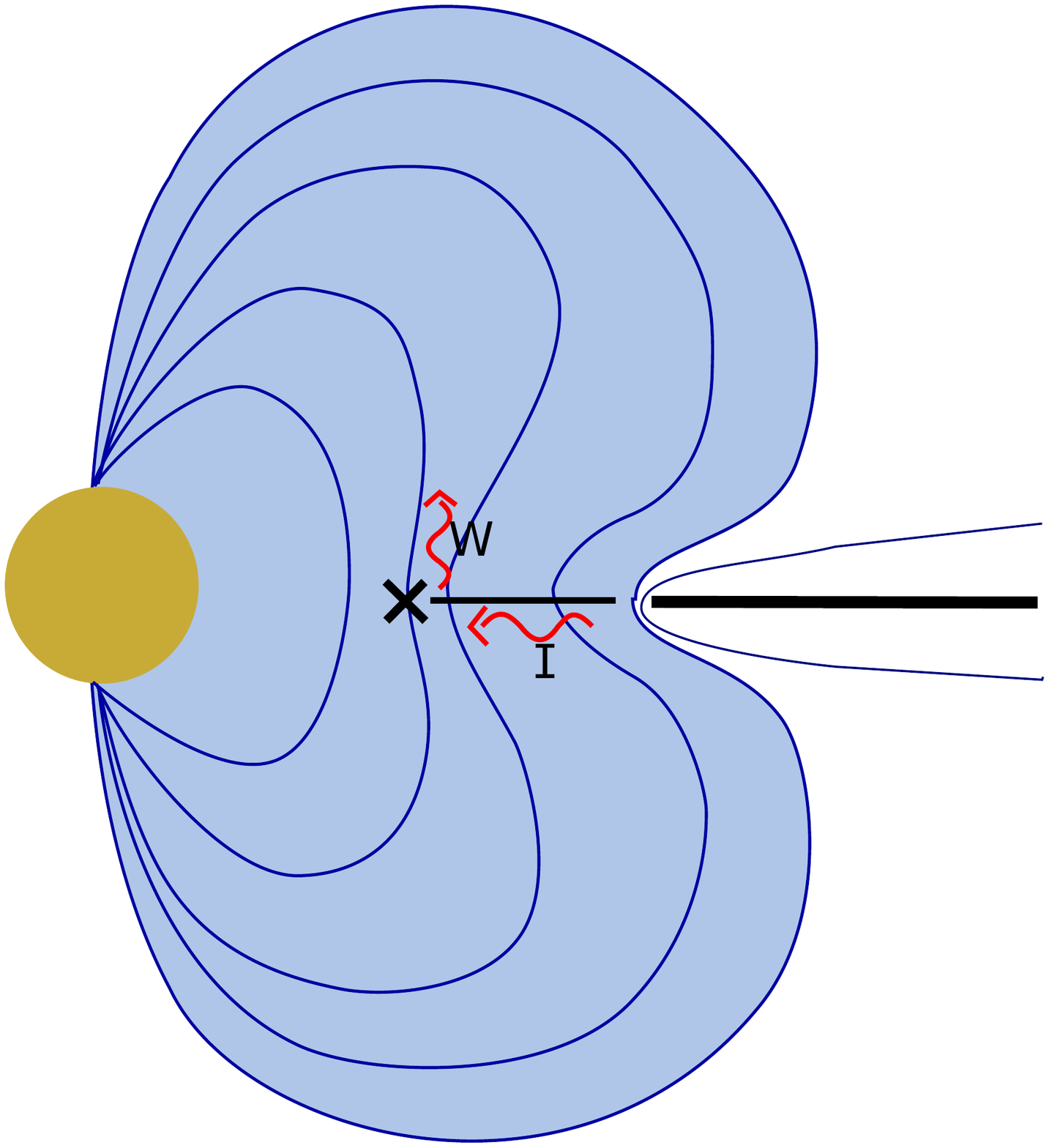}
 \caption{Schematic representation of the model. The blue region is the neutron star magnetosphere, pinched by the keplerian accretion disc. The magnetospheric disc  inside the magnetosphere is supported against gravity by magnetic tension, while the keplerian disc  is a usual, centrifugally supported, accretion disc. The arrows schematise the physical process involved in the model: as the gas reaches the inner edge of the keplerian disc it is picked up by magnetospheric field lines and slows down to the neutron star spin frequency $\nu_*$. Supported against gravity by magnetic tension, it is subject to the interchange instability which transports it inward toward the magnetic saddle point (marked by an X in the figure) where the concavity of the field lines changes. Radiation from the keplerian disc and the stellar surface then makes it unstable to a warp wave, which can trigger magnetospheric accretion vertically along the field lines.} \label{fig:schema}
\end{figure}
%
\section{Magnetospheric disc} \label{magnetospheric}
\subsection{Magnetic geometry}\label{geo}
Let us first describe the geometry we consider, as shown in figure \ref{fig:schema}: the keplerian accretion disc presses on the dipole magnetic field of the neutron star, so that a fraction of the magnetospheric field lines are pinched and have a concavity looking outward. There is necessarily a boundary layer \citep[see][]{GL79} where plasma from the accretion disc is picked up by magnetospheric field lines, rapidly slowing down from the keplerian rotation frequency (typically of the order of 1 kHz) to the neutron star spin frequency (typically a few times slower). As previous authors \citep{SPR90,LEP96}, and contrary to \cite{GL79}, we assume that this boundary layer is very thin. For this and a general discussion of the physics of the interaction between the disc and the stellar dipole field we will refer to the recent works of \cite{ROM03} and \cite{BZ08}, where the emphasis was on T Tauri stars but where the general setup is similar.\\
 A strong toroidal current must reside in this boundary layer, marking the transition from field lines anchored on the neutron star to ones anchored in the disc, and the radial Lorentz force from this current must, at equilibrium, balance self-consistently gravity and centrifugal force in the transition from neutron star to keplerian rotation frequency. It is important to note here that a large fraction of the luminosity from the system is emitted in this boundary layer, where the gas is rapidly slowed down from the keplerian frequency at the inner edge of the disc to the neutron star rotation frequency, and at the surface of the neutron star.\\
 In the resulting geometry there must thus be within the magnetosphere, as shown in figure \ref{fig:schema}, a radius where the concavity of the field lines changes from inward to outward. Using cylindrical coordinates $(r,\phi,z)$, it is easy to show that at this radius and in the disc midplane, the vertical component of the magnetic field $B_z$ starts increasing with radius: indeed, following a field line upward from the midplane (where $B_r=0$), one sees that $B_r$ is negative on a field line within this radius, and positive outside this radius: thus $\partial B_r/\partial z$ changes sign at this radius and so does $\partial B_z/\partial r$, since outside the midplane we assume to be in vacuum with no currents, giving $\vec \nabla\times\vec B=0$. We call this region, marked by an X in figure \ref{fig:schema}, the magnetic saddle, since one easily shows that $\partial^2|B|/\partial z^2$ is negative while $\partial^2|B|/\partial r^2$ is positive ; we call the corresponding radius $r_{saddle}$. We do not expect that force-free currents, which could exist if the magnetosphere is filled by a low-density plasma, would change this geometrical property.
 \subsection{Interchange instability}
 \cite{SPR90, LEP96} and \cite{AGA97} have studied the MHD equilibrium and stability of the magnetospheric disc formed of plasma trapped in this configuration. One instability involves essentially radial motion, and the other one vertical motion. The first one is an interchange instability, classical in plasmas supported by magnetic fields against gravity: it interchanges flux tubes and the plasma they contain, and thus releases gravitational energy, if more plasma moves radially in than out, \ie~if initially the mass per unit magnetic flux  increases outward.  \cite{SPR90} showed that this could allow plasma from the accretion disc to penetrate in  the magnetosphere, and this gas forms what we call the magnetospheric disc, suspended against gravity mostly by the magnetic field;  \cite{LEP96} derived the stability condition of the interchange mode in the magnetospheric disc as (in our notations):
 \begin{equation}
 \frac{\partial}{\partial r}\left(\frac{\Sigma}{|B_z|}\right)\le\frac{r|B_z|}{2\pi G M\Lambda}
 \label{eq:inter}
 \end{equation}
where $\Sigma$ is the surface density in the disc (assumed infinitely thin), $M$ is the mass of the neutron star, and $\Lambda$ is a parameter describing the geometry of the field line.\\
 In this equation the left-hand side is the radial derivative of the mass per unit of magnetic flux, as expected from the physics of the interchange instability. The right-hand side describes the stabilising effect of line tying, \ie~Êthe fact that the field lines are anchored on the star: this implies that, although the instability involves motion essentially transverse to the field lines, some torsion must be applied to them as they are exchanged at the disc but not at the star surface. The energy this requires decreases or cancels the gravitational energy released in the interchange, stabilising it. $\Lambda$ is infinite on open field lines, is large on the outermost magnetospheric field lines, because they are long and can be twisted easily, and becomes $\sim$Ê1 and strongly stabilising on the innermost field lines. The presence of $\Sigma$ in the left-hand side implies that the stabilizing effect is extremely strong if the surface density of the magnetospheric disc is low, \ie~the release of gravitational energy cannot compete with the necessary magnetic energy.\\
Even without line-tying one can show that the interchange, near marginal stability, cannot transport the gas inward from the magnetic saddle, which is also the radius where $\Lambda$ decreases sharply. We thus conclude that the interchange instability results in penetration of the gas from the keplerian disc, forming a tenuous disc where line tying is not yet efficient and accumulating where line tying becomes strong and quenches the instability. The exact radius where this happens cannot be farther in than the magnetic saddle, and depends very strongly on details of the keplerian disc/magnetosphere interaction, which could be obtained only from a self-consistent computation such as the ones of \cite {BZ08}, in 3 (rather than 2.5) dimensions in order to show the interchange mode. \\
This (a sudden drop in density at the magnetospheric radius, and a local density maximum inside the magnetosphere) is indeed observed in the simulations of \cite{BZ08}, and shown in their figure 4. However there the penetration of the gas within the magnetosphere is due to dissipation and can thus only roughly represent the action of the interchange mode, whose development is prevented by the assumed axisymmetry. The quenching of radial transport, resulting from line tying stabilisation of the interchange mode, is thus not taken into account.
\section{Warping instability}\label{warping}
\subsection{Bending wave}
What happens beyond this was tentatively described by \cite{SPR90} as resulting from the vertical MHD instability, which causes a warp of the magnetospheric disc and gives the gas some vertical motion: it  can thus be expected to trigger  magnetospheric accretion, vertically along the field lines. However \cite{LEP96} and \cite{AGA97} refined the theory of this instability, taking into account rotation. 
We follow their simple approach (an aligned rotator), where the disc is infinitely thin and the neutron star magnetic axis is aligned with its spin, so that the unperturbed system is axisymmetric and the magnetic field purely poloidal: $\mathbf{B}=(B_r,0,B_z)$.\\
Using a WKB approximation for the radial structure of the bending wave, \citet{AGA97} find its dispersion relation:
\begin{equation}
(\nu_v-m\nu_*)^2 = \nu_K^2+\frac{B_r^+}{2 \pi^2 \Sigma}\frac{\partial B_z}{\partial r}+\frac{2(B_r^+)^2}{4 \pi ^2 \Sigma}|k|
\label{eq_dispersion0}
\end{equation}
where $\nu_v$ is the frequency of the wave, $\nu_*$ is the neutron star spin frequency, $\nu_{K}$ the vertical epicyclic frequency (equal to the keplerian rotation frequency), $B_r^+$ is the radial magnetic field at the disc surface, $\Sigma$ the disc  surface density and k and m are the radial and azimuthal wavenumbers of the perturbation $\xi_z\sim\exp[i(kr+m\phi)-2 i\pi \nu t]$.\\
In this equation, the right hand side terms are given by the vertical restoring forces applied to the disc when it is displaced from its equilibrium position: the gravitational force appears through the epicyclic frequency, and the Lorentz force through the last two terms which correspond respectively to magnetic tension and magnetic pressure.\\
Furthermore, in equation \ref{eq_dispersion0}, $B_r^+$ is due to the azimuthal current in the disc, and this current in turn is responsible for the radial support of the gas, suspended in the gravitational field by magnetic tension: the radial equilibrium is maintained by the balance of  the verticaly integrated gravity, centrifugal and Lorentz forces:
\begin{equation}
\Sigma \frac{\partial \Phi}{\partial r}-4\pi^2\nu^2_*r \Sigma-2B^+_r B_z=0
\end{equation}
where $\Phi$ is the gravitational potential of the neutron star, linked with the keplerian frequency, so that one finally gets \citep{AGA97}
\begin{equation}
 \frac{B^+_r}{\Sigma}=\frac{4 \pi^2r(\nu^2_K-\nu^2_*)}{B_z}
 \end{equation}
Thus equation \ref{eq_dispersion0} can be rewritten as
\begin{eqnarray}
(\nu_v-m\nu_*)^2 = \nu_K^2&+&\frac{r(\nu_K^2-\nu_*^2)}{B_z}\frac{\partial B_z}
{\partial r}\nonumber\\
&&+\ 8\pi^2r^2(\nu_K^2-\nu_*^2)^2\frac{\Sigma}{B_{z}^2}|k|
\label{eq_dispersion}
\end{eqnarray}
The surface density of the disc thus disappears from the second term in the right-hand side, which now depends only on the magnetic configuration. The third term, on the other hand, can in a first approach be neglected if the surface density of the gas is low enough.\\

In this approach one gets the stability criterion, for the MHD warp wave, from the condition that the right-hand side of equation \ref{eq_dispersion} be positive:
\begin{equation}
\frac{r}{B_z} \frac{\partial B_z}{\partial r}+\frac{\nu_K^2}{\nu_K^2-\nu_*^{2}}\ > \ 0
\label{eq:vert}
\end{equation}

Since observations show that at the inner edge of the keplerian disc $\nu_K$ is typically 2 to 4 times higher than $\nu_*$, and is even higher in the magnetosphere, the second term is strongly stabilising so  that the vertical mode becomes active only at a significant distance inward from the magnetic saddle (where $\partial B_z/\partial r=0$). This leaves a radial gap where the interchange can no more cause the gas to move in, and the vertical instability cannot yet cause it to move up along the field lines: the conventional MHD instabilities thus still leave a missing link between radial and magnetospheric accretion.\\
\cite{ROM02}~and \cite{BZ08} find in their simulations that  funnel flows can be initiated by the vertical pressure gradient in the gas. We note however that in this region of a neutron star magnetosphere (in contrast with the protostellar case with which these works are more concerned) the gas should be submitted to very rapid cooling, whereas in the absence of differential rotation, and assuming that the interchange mode is stabilised by line tying, there is no known source of turbulence to heat it, so that the formation of a pressure-driven funnel flow may be difficult.\\
In the present work we thus return to Spruit's original idea and present a model that could provide  an alternative trigger for magnetospheric accretion. For this we will consider an additional mechanism that can make the warp unstable. Interestingly, this mechanism relies on the intense radiation field in the vicinity of an accreting neutron star. 
\subsection{When the wave becomes an instability}\label{instab}
%
	We now turn to the radiative mechanism that can make the bending wave unstable, permitting it to reach high amplitudes. For this we use a variant of the model presented by \cite{PRI96} to explain the warp observed in the outer region of the accretion disc of AGNs. In that case radiation coming from the {\it inner} region of the disc and pressing on the surface of the {\it outer} disc region was shown to make the warp wave unstable. Here on the other hand we are interested in an instability of the innermost (magnetospheric) region of the disc, and radiation can be assumed to come {both from the surface of the neutron star and} from the inner region of the keplerian disc, immediately outside the magnetospheric disc. We find that {both} can result in warp instability, and we present here a derivation only for illumination from the keplerian disc (\ie~from outside) since illumination from the stellar surface (\ie~from inside) is essentially the case studied by Pringle -- with the difference that our disc is in solid rather than keplerian rotation, at the stellar spin frequency $\nu_*$.\\
We assume for simplicity that  the radiative flux is axisymmetric and comes from a transition ring at the magnetospheric radius; integrating over angles we find that the luminosity  dL pressing on a surface dS of the magnetospheric disc is, in the same polar grid as before:
\begin{equation}
dL=\frac{L}{4\pi x^2}|\mathbf{u_x.dS}|
\end{equation}
where x is the radial distance between the surface element dS and the magnetospheric radius. Then the pressure force is
\begin{eqnarray*}
dF&=&\frac{2}{3c}dL\\
\frac{dF_z}{dS}&=&\frac{L}{4\pi x^2} \frac{2}{3c}\frac{\mathbf{dS}}{dS}.\frac{\mathbf{x}}{x}\\
&=&\frac{L}{6\pi c x^3}(-\partial_r\xi_z,-\frac{1}{r}\partial_\theta \xi_z,1).(-x,0,\xi_z)\\
&=&\frac{L}{6\pi c x^2}(\xi_z/x+\partial_r\xi_z)
\end{eqnarray*}
where c is the velocity of light and $\xi _z$ the vertical displacement of the surface element.
Adding this force in the equations of motion we get the dispersion relation, modified from equation \ref{eq_dispersion}:
\begin{eqnarray}
(\nu_v-m\nu_*)^2=\nu_K^2+2\frac{r(\nu_K^2-\nu_*^2)}{B_z}\partial_rB_z+\frac{B_r^2}{2\pi^2\Sigma}\vert k\vert \nonumber\\
-\frac{L}{6\pi c x^3 \Sigma}-i\frac{L}{6\pi c x^2\Sigma}\ k
\label{eq_fulldisp}
\end{eqnarray}
Taking for definiteness $m$ positive, let us first consider the root 
\begin{eqnarray}\label{prograde}
\nu_v-m\nu_*=\bigg[\nu_K^2+2\frac{r(\nu_K^2-\nu_*^2)}{B_z}\partial_rB_z+\frac{B_r^2}{2\pi^2\Sigma}\vert k\vert & \nonumber\\  
 -\frac{L}{6\pi c x^3 \Sigma}-i\frac{L}{6\pi c x^2\Sigma}\ k&\bigg]^{1/2}
\end{eqnarray}
To lowest order, the wave frequency will be close to
\[
\nu_v\approx m\nu_*+\nu_K.
\]
Treating the luminosity contributions as perturbations in order to identify the trend of their consequences,  one finds easily that
\begin{itemize}
\item waves with negative $k$, \ie~{\it leading} waves are amplified by radiation (as in Pringle's mechanism)
\item these waves propagate radially inward, from the keplerian disc/magnetospheric disc interface toward the magnetic saddle
\item their radial group velocity is proportional to $\Sigma$, \ie~they propagate very slowly if the surface density of the magnetospheric disc is weak. This leaves them ample time to be amplified from small perturbations at the disc/magnetosphere interface.
\end{itemize}
As in Pringle's mechanism, waves with a large radial wavenumber have stronger linear amplification, but we expect that, as in AGN discs, non-linear and viscous effects (see \cite{OGI01} and references therein) will select  the dominant mode. We note that in AGN the dominant observed mode has $m=1$ and a small $k$ (small winding angle), but there as in neutron star binaries modes of smaller wavelength might result in a weaker modulation and be more difficult to observe.\\
Taking $k\sim 1/x$ as the smallest possible wavelength, we note that the real and imaginary contributions due to the radiation field are of the same order in equation \ref{eq_fulldisp}. This implies that a luminosity sufficient to induce strong amplification will also significantly decrease the real part of the frequency. %
\subsection{The retrograde wave}
The dispersion relation, equation \ref{eq_fulldisp}, also admits a retrograde wave obtained by taking the negative square root in the right-hand side:
\begin{eqnarray*}
\nu_v-m\nu_*=-\bigg[\nu_K^2+2\frac{r(\nu_K^2-\nu_*^2)}{B_z}\partial_rB_z&+&\frac{B_r^2}{2\pi^2\Sigma}\vert k\vert \nonumber\\  
-\frac{L}{6\pi c x^3 \Sigma}&-&i\frac{L}{6\pi c x^2\Sigma}k\ \bigg]^{1/2}
\end{eqnarray*}
Examination of this relation shows that this wave would have the same characteristics as the ``direct'' one we have discussed: leading waves with negative $k$ are amplified and propagate inward  with the same growth rate and group velocity. However we note that a retrograde wave would also be possible in Pringle's model for AGNs, but does not seem to appear in observations. We believe that this might be due to the very low level, at these frequencies, of the thermal noise from which the unstable wave has to grow. We will thus ignore the retrograde wave in  our discussion.
\section{Towards a mechanism for kHz Quasi-Periodic Oscillations ?} 
If the radiation field is sufficient to make the warp wave strongly unstable, one should expect this to have observable consequences since it would result in the presence of a tilted surface, rotating at a frequency in the kHz range in the immediate vicinity of the most emissive regions of the star-disc system. Both obscuration of emissive regions and reflection of the luminosity should result in a strong modulation of the X-ray signal at that frequency, \ie~in a Quasi-Periodic Oscillation (QPO). In this section we explore the possibility that the radiation-driven warp could explain one of the twin kilohertz QPOs observed in low-mass X-ray binaries that host a neutron star. We find that, although this could solve some questions  raised by models of these QPOs, it results in frequencies that do not correspond to the observed ones. This estimate is however still limited by the available models of the disc-magnetosphere interface. We will finally discuss possible properties of the disc-magnetosphere interface, not included in these models, that could produce warp frequencies in better agreement with the observed QPOs. 
\subsection{Observations}
The frequencies $\nu_1$ and $\nu_2$ of the lower and upper kHz QPO can change by a factor of two in a given source but their separation $\Delta \nu=\nu_2-\nu_1$ varies little and was initially observed to stay close to the frequency of the burst oscillations, believed to be the neutron star spin frequency $\nu_*$.  However $\Delta \nu$ is observed to decrease by a  significant fraction when $\nu_1$ and $\nu_2$ reach their highest values. Until recently it was generally thought that  { $\Delta \nu$ stayed close to either the burst oscillation frequency or its half}, but after a new examination of the data \citet{MEN07} have shown that the link between the two quantities is not so direct, as we will discuss below.\\

\subsection{A new approach}\label{approach}

Numerous models have been proposed to explain the different types of QPO in
LMXB. For the neutron star binaries kHz QPO, two main classes of models can be
distinguished: those based on a beat frequency and those based on relativistic
precession motion. (\cite{LAM85}, \cite{MIL98}, \cite{STE98}, \cite{STE99})\\
We note here that a presupposition of both of these models, that $\nu_{2}$ corresponds to the keplerian rotation frequency at the inner disc edge, has remained widespread without further justification. 

In these interpretations of the kHz QPO, two points stand out. The first one is that in these models the most `fundamental' mode is the one 
associated with the keplerian rotation frequency at the inner disc edge, at frequency $\nu_{2}$. We would thus expect the higher kHz QPO, to
have a higher coherence, and thus a higher $Q$ factor (where $Q=\nu / FWHM$, and  $FWHM$ is the frequency width at half maximum) 
than the lower one at frequency $\nu_{1}$. But the opposite is observed, \ie~the lower kHz QPO can have an extremely high $Q$, much 
higher than the other QPO (see \eg~\citet{BAR05}). 

 	The second point we wish to address has been emphasised by \citet{VDK00}: although the beat frequency model fails to explain the variations of \dn, the proximity between \dn~and $\nu_*$ would indicate that somehow the gas disc seems to `know' about the neutron star spin frequency. This would be very difficult to explain,  {since any direct connection (\eg~by magnetic field lines) between the disc and the neutron star or its magnetosphere would mechanically imply a constant \dn}.\\
In the present work we consider the gas that has entered the magnetosphere, and thus rotates at $\nu_*$ (and will presumably end up accreted along magnetospheric field lines). Mirroring the previous argument, it would be difficult for this gas to know about the keplerian rotation frequency in the disc. However this gas feels the same gravity field that corresponds to the keplerian rotation, although magnetic support allows it to rotate solidly at the stellar  frequency. This gravity field also defines the vertical epicyclic frequency of the magnetospheric gas, so that according to equation \ref{prograde} the warp frequency $\nu_{v}$ should be close to $\nu_{K}+\nu_{*}$ for an $m=1$ mode. Thus assuming that the lower kHz QPO (rather than the higher one as in other models) appears at the rotation frequency at the inner disc edge, one could expect the warp to appear as a higher-frequency QPO with a frequency difference close to $\nu_{*}$. \\
However additional terms appear in equation \ref{prograde}, and can substantially change the warp frequency. The main unknown quantities that can affect the frequency difference \dn~are
\begin{enumerate}
\item the effect of the second term, due to the gradient of $B_{z}$, in the right-hand side of equation \ref{prograde}.
\item the distance between the magnetospheric radius, where we assume that the QPO at frequency $\nu_1$ originates, and the radius where  the frequency of the warp must be computed
\end{enumerate}
This should be obtained from a self-consistent computation of the magnetospheric disc equilibrium, in three dimensions and with low enough dissipation to show the interchange instability and its consequences on the accretion of the gas. This  is far beyond present theoretical or numerical possibilities.  \cite{LEP96} have studied simple exact models of such equilibrium configurations, in the case of a partially diamagnetic keplerian disc. They found a class of stable equilibria that we can use to estimate representative values of the magnetic terms of the dispersion relation. We present in appendix detailed calculations of these terms, which let us calculate the resulting warp frequencies using equation \ref{prograde}. \\
Fig \ref{fig2}a shows the resulting evolution of $\Delta \nu$ as a function of  $\nu_*$ in one of these models. The free parameters are the neutron star mass, spin and magnetic moment, the position $r_{M}$ and $r_{i}$ of the inner edges of the keplerian and magnetospheric disc, and finally the radius $r_0$ of the magnetospheric disc where the frequency of the wave is computed. We have chosen the units such as $r_{M}=1$, $GM=1$ where $G$ is the gravitational constant and $M$ the neutron star mass. The magnetospheric disc extends between $r_i=.8$ and $r_{M}=1$, and the frequency of the wave is computed at $r_{0}=.8 $, the location of the magnetic saddle.\\ 
The general trend is that, as observed when comparing different sources, $\Delta \nu$ decreases when $\nu_*$ increases, although the ratio between them is higher than observed, but this depends both on the radii chosen and on the choice we have made in the class of equilibria of \cite{LEP96}. 
\subsection{The disc/magnetosphere interface}\label{interface}
\begin{figure}
   \centering
\includegraphics[width=0.5\textwidth]{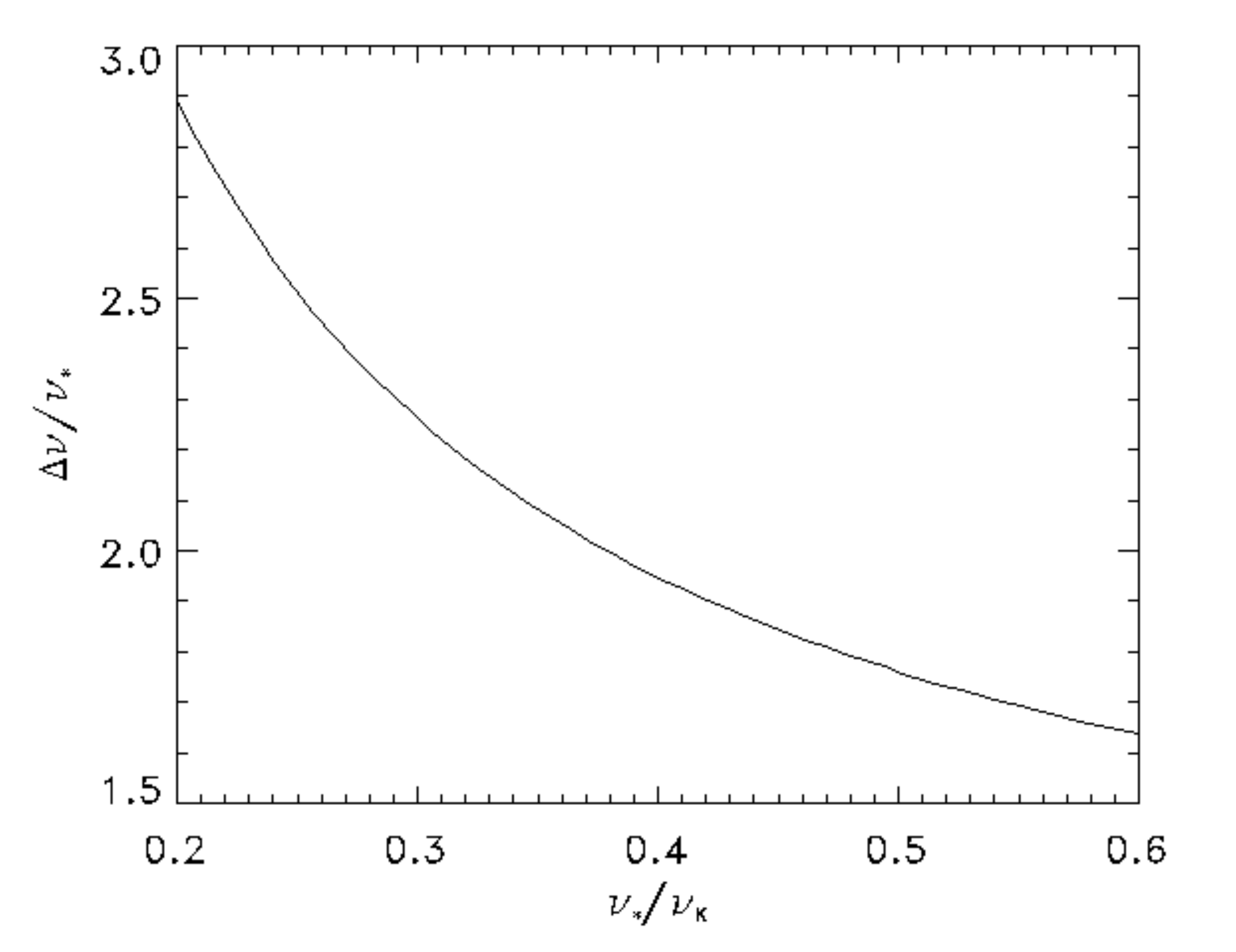}
\includegraphics[width=0.5\textwidth]{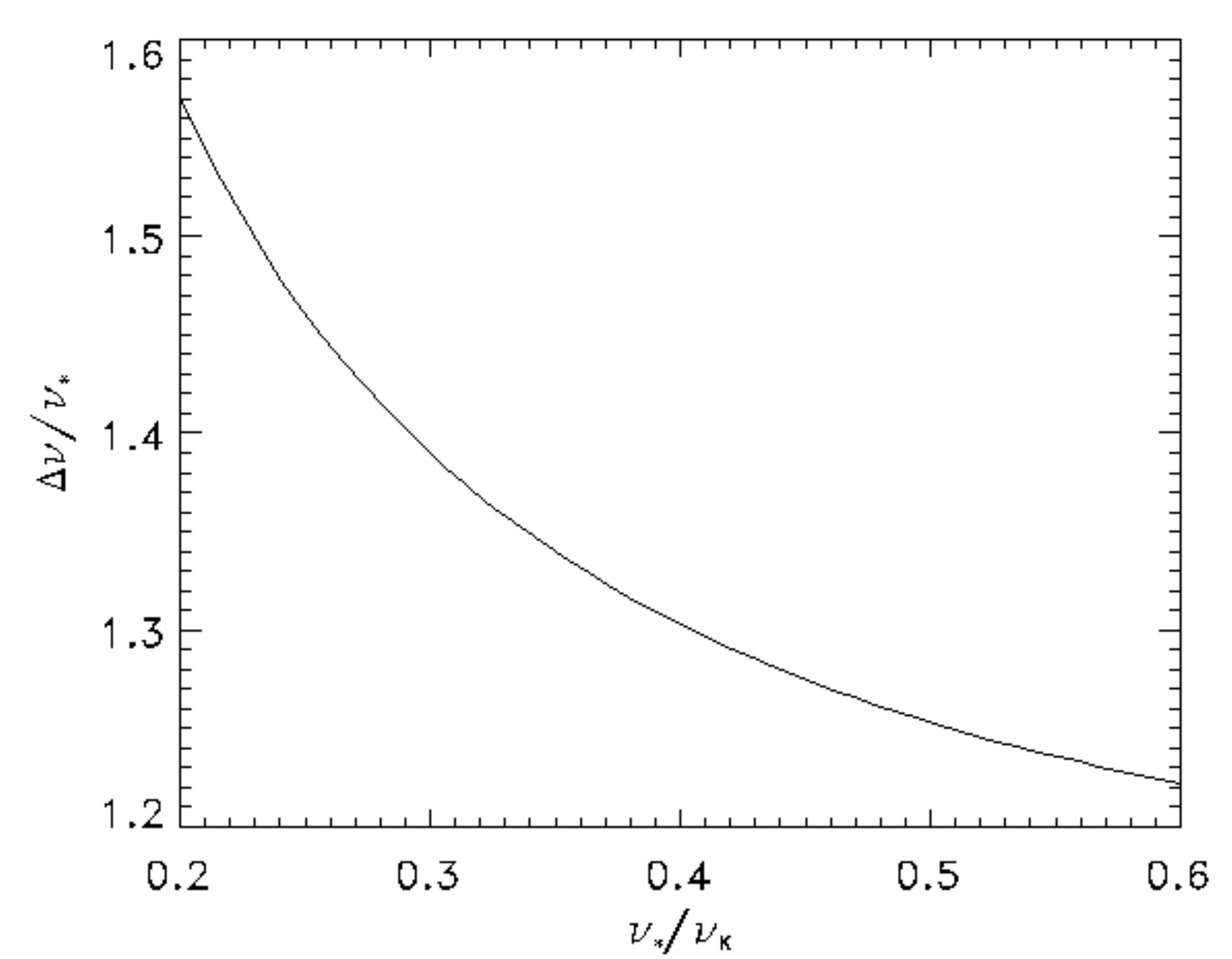}
\includegraphics[width=0.5\textwidth]{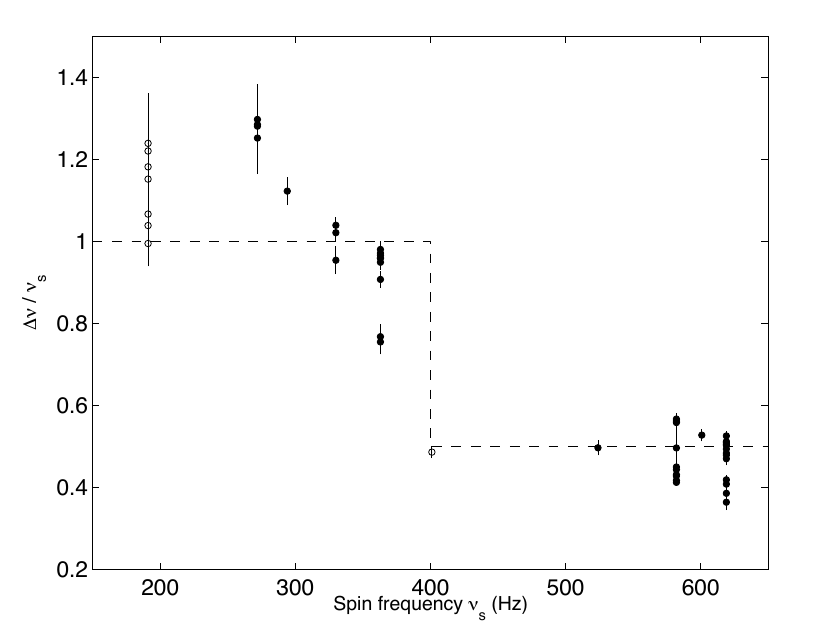}
 \caption{The upper plot shows the dimensionless quantity 
${\nu_v(r_0=0.8r_{M})-\nu_K(r_{M})}/{\nu_*}={\Delta \nu}/{\nu_*}$ in our model, 
when ${\nu_*}/{\nu_K(r_{M})}$ varies. $\Delta \nu$ is the frequency difference
between the two kHz QPO, $\nu_*$ is the neutron star spin frequency and
$\nu_K(r_{M})$ is the rotation frequency at the inner edge of the keplerian
disc. The lower plot is the observational data as plotted by Mendez and
Belloni (2007). 
The shape of the curves is similar, although the numerical values differ, as explained in the text.}
 \label{fig2}
   \end{figure}

\begin{figure}
   \centering
	\includegraphics[width=0.5\textwidth]{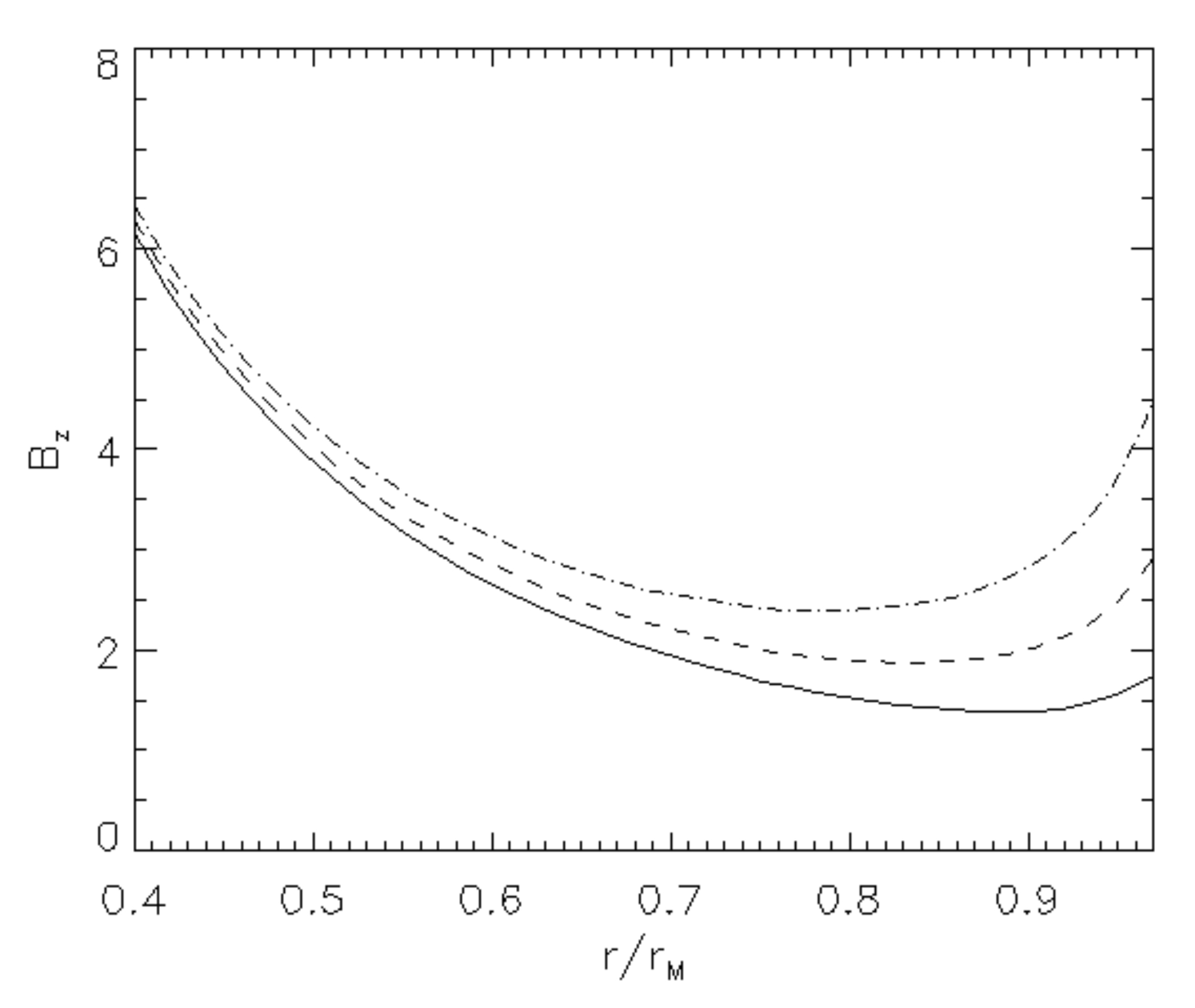}

 \caption{ Vertical component of the magnetic field in the plane of the disc in function of radius. The straight, dash and dotted dash lines correspond to different values of the diamagnetic parameter: respectively $\lambda=0.3, 0.6, 1$}
 \label{Bz}
   \end{figure}

Figure \ref{fig2}b shows the same plot as Figure \ref{fig2}a using a different model, where the saddle is closer (at $r_{0}=.88$) and the disc is not fully diamagnetic, so that it is threaded by a fair amount of the neutron star magnetic flux. \dn~ is decreased but still remains larger than $\nu_{*}$. Obviously at this stage these results do not permit to consider the radiatively-driven warp as an explanation for the observed QPO. There are however two effects that could reduce \dn~Êto values compatible with the observations. \\
The first effect is the role of the luminosity, in equation \ref{prograde}. It contributes both to the growth rate and to the real part of the frequency, in a ratio equal to $kx$. Let us thus consider a luminosity high enough to make the warp strongly unstable, with a growth rate of the order of the rotation frequency: its effect on the warp frequency, for a large-scale warp ($kx\sim 1$) will thus be to decrease \dn~Êby a similar amount, bringing it to a better possible agreement with the observations - in particular the fact that \dn~decreases at high luminosity.\\
The second effect can be understood by considering the radial profile of $B_{z}$, as shown in figure \ref{Bz} with different sets of parameters: $B_{z}$ first decreases outward (at low radius the dipole field of the neutron star is weakly affected by the disc), then has a minimum (the magnetic saddle) and increases toward the magnetospheric radius: this is due to the fact that in the basic model the disc is fully diamagnetic, rejecting the neutron star magnetic field: one thus has a strong current ring at the magnetospheric radius, explaining the sharp rise of $B_{z}$ which strongly increases the warp frequency, according to equation \ref{prograde}. When the diamagnetism (the parameter $\lambda$ of \cite{LEP96}) is reduced, a part of the stellar field threads the disc, reducing the ring current and the rise of $B_{z}$ - and finally \dn, as in figure \ref{fig2}b.\\
Such a reduction of the diamagnetism is probably unrealistic, since the stellar field lines rotating at $\nu_{*}$ in the disc would involve an extremely strong, and probably unrealistic, turbulent resistivity. However the model doesn't include a magnetic field originating in the disc itself, with magnetic field lines open to infinity rather than tied to the star. We believe that the inclusion of this magnetic contribution would also reduce the ring current without requiring resistivity, and might result in a warp frequency in better agreement with the observed QPO frequency.
%
%
\section{Conclusion}
In this work we have explored a variant of the radiation-driven instability, found  by \cite{PRI96} to explain the warps observed in the outer region of AGN discs. In this variant radiation from the inner region of a neutron-star accretion disc destabilizes the gas that has entered the  stellar magnetosphere and formed a magnetospheric disc. The mechanism is similar, but can be expected to be even more effective as the radiating region and the destabilized one are in very close contact.\\
The resulting warp wave can be expected to play an important role in the chain that permits the gas, first accreted in a conventional manner in the keplerian disc, to enter the magnetosphere by the interchange instability as found by \cite{SPR90}, to form a stable magnetospheric disc as shown by \cite{LEP96}, and eventually to accrete onto the neutron star by flowing along its magnetic field lines. More specifically, we have discussed how the warp could provide a crucial missing link between the interchange and magnetospheric accretion.\\
We have finally discussed how this warp, forming a non-axisymmetric feature in the most strongly emitting region of the star-disc system, should produce a strong modulation of the luminosity by reflexion and obscuration of the accretion power.  This could produce a Quasi-Periodic Oscillation in the kHz range but, at the present stage of models of the magnetosphere/accretion disc interface, its frequency is not adequate to reproduce the observed QPO frequency. \\
On the other hand the warp would be well adapted to explain some other properties of the kHz QPOs, such as the variation of he frequency with the luminosity. This will be an incentive for further work that will aim at perfecting the models of the magnetosphere/accretion disc interface, in order to explore possibilities that could result in warp frequencies in better agreement with QPO observations.

\section*{Acknowledgments}
The authors gratefully acknowledge very helpful discussions with D. Barret, M. van der Klis, F. Casse and P. Varniere. 
\appendix
\section[]{Magnetic configuration}
\begin{figure}
   \centering
	\includegraphics[width=0.5\textwidth]{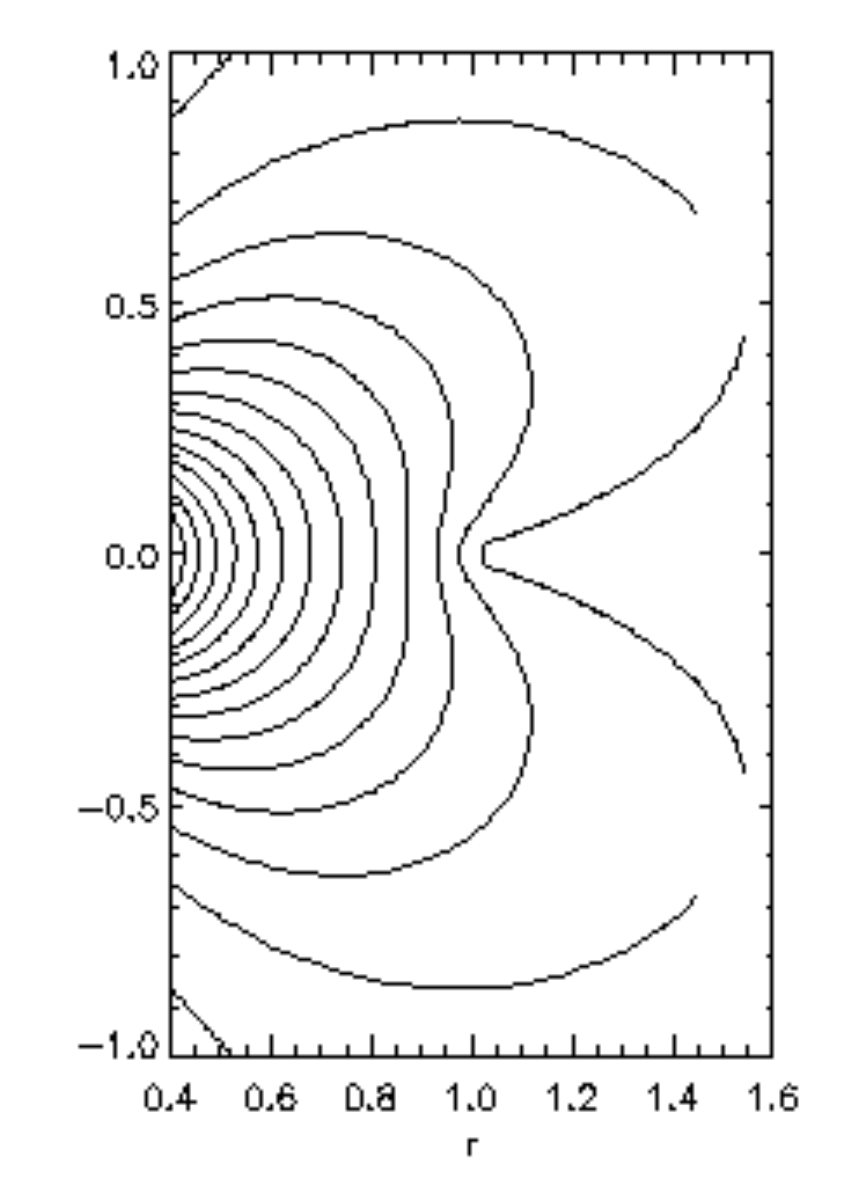}
      \caption{The disc-magnetosphere configuration computed in the appendix.}
       \label{fig4}
   \end{figure}
%
We present here the model used for the calculation of the magnetic terms of the dispersion relation, equation \ref{eq_fulldisp}, using the equilibria of \cite{LEP96} which describe a neutron star surrounded by a keplerian disc and a magnetospheric disc. They show that the magnetic field can written as 
\[\mathbf{B}=\mathbf{B_*}+ \mathbf{B_K}+\mathbf{B_M}\]
where the right-hand side contains contributions from respectively the stellar dipole field, currents in the keplerian disc and in the magnetospheric disc. Introducing a parameter $\lambda$ measuring  the diamagnetism of the keplerian disc, which excludes totally ($\lambda=1$) or partially ($\lambda<1$) the magnetic field from the neutron star, they find a family of equilibria where these contributions can be written as
\begin{eqnarray}
\mathbf{B_{*}}&=&\mathbf{B_1}((1-\lambda)\mu_1,\infty)\\
\mathbf{B_{K}}&=&\mathbf{B_1}(\lambda \mu_1,r_{M})\\
\mathbf{B_{D}}&=&\mathbf{B_3}(j,r_{M})
\end{eqnarray}
where $r_{M}$ is the inner radius of the keplerian disc and $j$ the azimuthal current in the magnetospheric disc. $\mathbf{B_1}(\mu,r_{M})$ and $\mathbf{B_3}(j,r_{M})$ can be derived from the azimuthal magnetic potentials $\mathbf{A}=f \mathbf u_{\phi}$ with (in a-dimensional units, convenient since for us only the magnetic geometry is important rather that the amplitude of the field)
\begin{eqnarray}
f_1(\mu,d)=\frac{2\mu}{\pi r}\Big(\frac{r^2X^2-d^2cos^2\theta}{d^2X}+sin^2\theta arctan X \Big)
\end{eqnarray}
where $X=\Big\{  \frac{1}{2}\Big(\big(1-\frac{d^2}{r^2}\big)^2+4 \frac{d^2}{r^2} cos^2\theta\Big)^{\frac{1}{2}}-\frac{1}{2}\Big(  1-\frac{d^2}{r^2}\Big) \Big\}^{\frac{1}{2}}$, and 
\begin{eqnarray}
f_3(j,d)=rsin\theta\int_0^\infty g(k) J_1(krsin\theta)e^{-krcos\theta}dk
\end{eqnarray}
where $J_1$ is the Bessel function of first kind and order 1, and 
\begin{eqnarray}
g(k)&=&\int_0^d\big( \frac{sin(ks)}{ks}-cos(ks) \big)\frac{G(s)}{s}ds\\
G(s)&=&\frac{4}{r_{i}^2}\int_0^s \frac{t^2j(t)}{\sqrt{s^2-t^2}}dt\\
j(r_{i}<r<r_{M})&=&j_0\frac{4r_{i}r_{M}}{(r_{M}-r_{i})^2}\frac{(r_{M}-r)(r-r_{i})}{r^2}
\end{eqnarray}
and c is the speed of light.
For our examples shown in figure \ref{fig2} we have chosen $\mu_1=10$, $j_0=-10$, $r_{i}=0.8$, $r_{M}=1.$, and respectively $\lambda=.8$ and $\lambda=.265$.
Fig \ref{fig4} gives the shape of the corresponding magnetic field lines for $\lambda=.8$.

\bsp
\label{lastpage}
\end{document}